\documentclass[letter]{aa}
\usepackage{natbib}
\usepackage{xcolor}
\usepackage{graphicx}

\begin{document}

\title{Understanding the white-light flare on 2012 March 9 : Evidence
    of a two-step magnetic reconnection}

\author{Q. Hao\inst{1} \and Y. Guo\inst{1,2} \and Y. Dai\inst{1,2} \and M. D. Ding\inst{1,2} \and Z. Li\inst{1,2} \and X. Y. Zhang\inst{3} \and C. Fang\inst{1,2}}

\institute{School of Astronomy and Space Science, Nanjing
University, Nanjing 210093, China \\ \email{fangc@nju.edu.cn}
\and
Key Laboratory of Modern Astronomy and Astrophysics
(Nanjing University), Ministry of Education, Nanjing 210093, China
\and
Yunnan Astronomical Observatory, Chinese Academy of Sciences, Kunming 620011,
China \\ \email{xyz@ynao.ac.cn}
}


\begin{abstract}
{}{We attempt to understand the white-light flare (WLF) that was observed on 2012 March 9 with a newly constructed
multi-wavelength solar telescope called the Optical and Near-infrared Solar Eruption Tracer (ONSET). }
{We analyzed WLF observations in radio, H$\alpha$, white-light,
ultraviolet, and X-ray bands. We also studied the magnetic configuration of the flare via the nonlinear force-free field (NLFFF) extrapolation and the vector magnetic field
observed by the Helioseismic and Magnetic Imager (HMI) on board the \textit{Solar
Dynamics Observatory} (\textit{SDO}).
}
{Continuum emission enhancement clearly appeared at the 3600 \AA\ and 4250 \AA\ bands,
with peak contrasts of 25\% and 12\%, respectively. The continuum emission enhancement closely
coincided with the impulsive increase in the hard X-ray emission and a
microwave type III burst at 03:40 UT. We find that the WLF appeared at one
end of either the sheared or twisted field lines or both. There was also a long-lasting phase in the H$\alpha$ and soft X-ray bands after the white-light emission peak. In particular, a second,
yet stronger, peak appeared at 03:56 UT in the microwave band.
}
{This event shows clear evidence that the white-light
emission was caused by energetic particles bombarding the lower solar atmosphere. A two-step magnetic reconnection scenario is proposed to explain the entire process of flare evolution, i.e., the first-step magnetic reconnection between the field lines that are highly sheared or twisted or both, and the second-step one in the current sheet, which is stretched by the erupting flux rope. The WLF is supposed to be triggered in the first-step magnetic reconnection at a relatively low altitude.}

\end{abstract}


\titlerunning{A white-light flare on 2012 March 9}
\authorrunning{Hao, Q. et al.}
\maketitle

\section{Introduction}
White-light flares (WLFs) are flares that have emission visible in
the optical continuum \citep{Neidig1993}. Research on WLFs can provide critical insight into the possible energy transport and heating mechanisms
in the lower solar atmosphere \citep{Neidig1989, Ding1999}. The first WLF was observed 153
years ago \citep{Carrington1859}. However, when including the first observations of WLFs from space \citep{Hudson1992},
up to now fewer than 150 WLFs have been reported in the literature. These are
thought to be rare and among the most energetic flaring events. In most
cases, enhanced continuum emission is detected in the Balmer and
Paschen continuum. In a few cases, however, enhanced emission in the
infrared continuum was also reported \citep{Liu&Ding2001, Xu2004}.

\cite{Hudson2006} used the white-light (WL) channel ($>$1500
\AA) on board the \textit{Transition Region and Coronal Explorer} \citep[\textit{TRACE};][]{Handy1999}
with a spatial resolution of $1^{\prime\prime}$ and detected WL emission for flares
down to \textit{GOES} class C1.6. They suggested that the WL continuum
may occur in all flares. \cite{Fletcher2007} also analyzed nine flares
observed by \textit{TRACE} in WL. However, because the \textit{TRACE} WL/ultraviolet(UV)
emission always has a higher contrast than that of the traditional WLFs and the observations are sometimes saturated, it is unclear whether the
\textit{TRACE} continuum is affected by UV emission. \cite{Jess2008}
used high resolution observations of the one-meter Swedish Solar
Telescope to detect WL emission in the blue continuum
around 3954~\AA \ with a peak intensity 300\% above the quiescent flux in a
C2.0 flare. This high emission contrast, although restricted to a small area,
is surprising and it remains to be verified whether it is a
common feature in WLF observations.

Using \textit{Yohkoh} observations, \cite{Matthews2003} reported 28
WLFs in the G-band and found that they have a strong association with hard X-ray
emission. \cite{Wang2009} studied 13 flares observed by
\textit{Hinode} \citep{Tsuneta2008} and found that there is a
correlation between the \textit{GOES} soft X-ray flux and the flare emission in the G-band.
However, as the author pointed out, since \textit{Hinode} G-band observations are
contaminated by CH band emission and the time resolution of the
observations is insufficiently high, it is unclear whether the peak values and the
cut-off visibility of WLFs are real.

Can the origin of WLFs be explained by heating and energy transportation in
the lower solar atmosphere, even around the minimum temperature region? Several mechanisms
have been proposed, including electron beam bombardment followed by
radiative backwarming \citep{Machado1989,Metcalf1990}, energy transportation by
Alfv\'en waves \citep{Fletcher2008}, magnetic reconnection in the
lower atmosphere \citep{Ding1999,Chen2001}, heating by chromospheric condensation
\citep{Gan1992} and so on. It is necessary to distinguish between two types
of WLFs \citep{Machado1986}. \cite{Fang1995} studied the different characteristics
of each one using both observations and atmospheric
non-local-thermodynamic-equilibrium modeling. In the case of type I WLFs, there exists a strong
time correlation between the peak of either the hard X-ray emission or the microwave
burst and the maximum of the continuum emission, there
is a strong Balmer jump in the spectra, and the Balmer lines, particularly the H$\alpha$ line, are usually strong and broad.
However, type II WLFs do not display the above features.
For type I WLFs, either the electron bombardment or the following backwarming or both would be plausible mechanisms \citep{Ding2003,Chen2005,Fletcher2007}; however, for type II WLFs, the mechanism is more likely to be magnetic reconnection in the lower atmosphere \citep{Chen2001,Jiang2010}.

In this letter, we report a WLF observed in radio, H$\alpha$, WL,
ultraviolet, and X-ray bands. We study in particular the magnetic configuration of the flare via the nonlinear force-free field (NLFFF) extrapolation and the vector magnetic field
observed by the Helioseismic and Magnetic Imager (HMI) on board the \textit{Solar
Dynamics Observatory} (\textit{SDO}). Our data analysis and results are presented in
Section~2, and both our discussion and conclusions are given in Section~3.

\section{Data analysis and results}

An M6.3 two-ribbon flare was observed at N15W01 in active region NOAA 11429
on 2012 March 9 by the Optical and Near-infrared Solar Eruption Tracer (ONSET),
which was constructed by Nanjing University in cooperation with
Yunnan Astronomical Observatory \citep{Fang2012}. The telescope ONSET is installed
at a new solar observing site (E$102.57^\circ$, N$24.38^\circ$) near the Fuxian Lake,
60 km far from Kunming. The seeing is both stable and good \citep{Liu&Beckers2001}.

The ONSET consists of four tubes: (1) a near-infrared vacuum tube, with
an aperture of 27.5 cm, operating at the He I line center and wings up to 10830$\pm2.5$ \AA \ with a
full width at half maximum (FWHM) of 0.25 \AA; (2) a chromospheric (H$\alpha$) vacuum tube, with
an aperture of 27.5 cm, working at the H$\alpha$ line center and wings up to $6562.8 \pm 1.5$ \AA \ with an FWHM of 0.25 \AA; (3) a WL vacuum tube, with an aperture of
20 cm, working in either the 3600~\AA \ or 4250~\AA \ continuum with an FWHM
of 15 \AA; and (4) a guiding tube.

The observations were made from about 03:00 UT to 05:00 UT at 3600
\AA, 4250 \AA, H$\alpha$ line center, and H$\alpha$ $6562.8 \pm 0.5$ \AA,
with a cadence of 1 min. The pixel size is about $1^{\prime\prime}$.
We also analyzed UV data at 1700 \AA\ and 1600 \AA, which were observed by
the Atmospheric Imaging Assembly (AIA; \citealt{Lemen2012}) on
board \textit{SDO}. Figure 1 depicts the images at 3600 \AA,
1700 \AA, and the H$\alpha$ line center before the WLF (03:25 UT),
at the peak time of the WLF (03:40 UT), and
after the WLF (03:55 UT). The WL emission can be clearly
seen in both the 3600~\AA \ and 4250~\AA \ continua at 03:40 UT. The size of the WLF
is about $3^{\prime\prime}\times3^{\prime\prime}$ with a duration of about 2-3 min. However,
even before the WLF peak time, the emission at both H$\alpha$ and 1700~\AA \
already appeared as bright threads at 03:25 UT. As seen in Figure 1, these threads were very close to
the polarity inversion line separating the two highly sheared ribbons. The
image at 1600~\AA \ is similar to that at 1700~\AA. After the WLF peak, the two flare ribbons seen in 1700~\AA \ and H$\alpha$ bands are separated more widely than before, which indicates
that the magnetic reconnection occurred continuously.

\begin{figure}
\centering
\includegraphics[width=0.4\textwidth,viewport=10 20 525 540,clip]{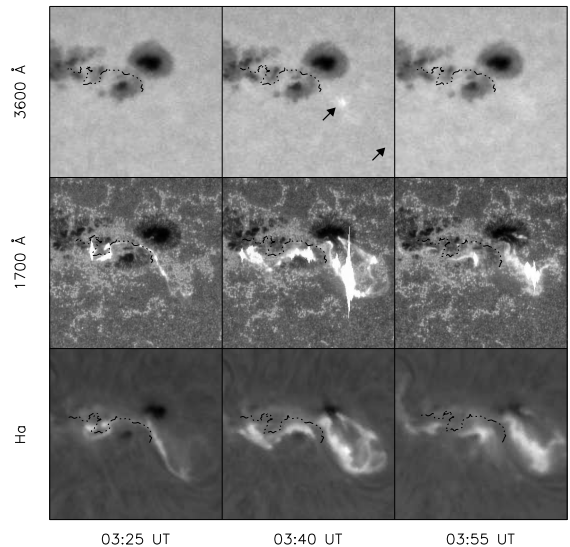}
\caption{Images showing the M6.3 flare evolution at 3600 \AA, 1700 \AA, and
the H$\alpha$ line center before the WLF (left column), at the peak time of the WLF (middle column),
and after the WLF (right column) on 2012 March 9. The field of view of each image is
$200^{\prime\prime}\times200^{\prime\prime}$. The location of the center of the field of
view is around ($30^{\prime\prime},380^{\prime\prime}$). The upper-left arrow indicates the area
where the white-light and H$\alpha$ intensities are measured as shown in Figure 2. The
lower-right arrow indicates the quiet region selected as the background. Dash-dotted lines
mark the polarity inversion line of the line-of-sight magnetogram.
} \label{fig1}
\end{figure}

\begin{figure}
\centering
\includegraphics[width=0.4\textwidth,viewport=10 0 560 680,clip]{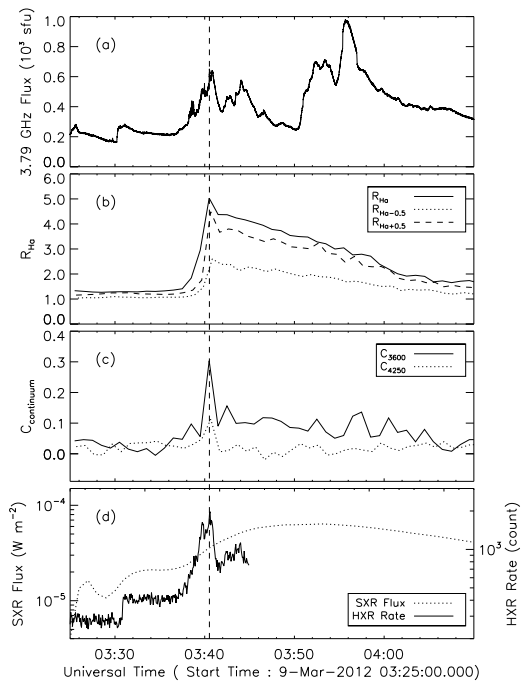}
\caption{Time profiles of the radio flux at 3.79 GHz, the intensity
enhancement $R=I/I_{0}$ at the H$\alpha$ line center and $6562.8 \pm 0.5$ \AA, the continuum
contrast $C=(I-I_{0})/I$ at 4250~\AA \ and 3600~\AA, the RHESSI hard X-ray (50--100 keV) flux,
and the GOES soft X-ray (1--8 \AA) flux of the WLF. The light curves in the
white-light and H$\alpha$ bands are computed in the WLF kernel as shown in Figure 1, and
those in other bands are computed for the entire solar disk.
The HXR data after 03:45 UT are omitted owing to a change of the various attenuators.
} \label{fig2}
\end{figure}

Figure 2 depicts the light curves of the WLF in different wavebands. We note that for (1) H$\alpha$,
$R=I/I_{0}$ and (2) both the 4250~\AA \ and 3600~\AA \ bands, the contrast is defined as $C=(I-I_{0})/I_{0}$, where
$I_{0}$ is the intensity of the nearby quiet region. It can be seen that
the peaks in different wavebands (except for the soft X-ray) are precisely consistent at around 03:40 UT,
which suggests that there is a close relationship between the energetic particles and the WL emission.
The microwave burst data were recorded by the Solar Radio Broadband Spectrometer (SRBS; Fu et al.
2004) at the Huairou Station of the National Astronomical Observatory of China. Near 03:40 UT, there is a type III burst with a positive frequency drift of 700 MHz s$^{-1}$ (Y. Y. Liu, private communication), which indicates that an electron beam bombarded the lower solar atmosphere from the upper atmosphere, which is supposed to be related to the observed WLF. In addition to the peak at around 03:40 UT, there is another series of peaks in the 3.79 GHz band, especially the main peak at 03:56 UT as shown in Figure~2(a). We checked other observations, such as extreme-ultraviolet (EUV) images, and found that there was enhanced emission
only in the active region NOAA 11429 from 03:40 UT to 03:56 UT. Therefore, the main peak at 3.79 GHz implies
that there were still energetic particles at 03:56 UT. Figure~2(b) shows that the H$\alpha$ line was strong
and broad both during the WLF peak and after it, since both the line center and line wing display strong
emission. Figure~2(c) shows that the contrasts at the WLF peak time in the 3600~\AA \ and 4250~\AA \
continua are about 25\% and 12\%, respectively. The peak time of the soft X-ray flux is about 03:53 UT, which is
approximately 13 min later than the hard X-ray peak time. This behavior is due to the Neupert effect, i.e.,
the thermal fluxes coinciding with the integral of the non-thermal fluxes. However, the 13 min delay is relatively
long \citep{Li2006} and the soft X-ray does not drop too much even 30 min after the hard X-ray peak.
This strongly suggests that the flare continuously released energy in phase after the WLF peak.

\begin{figure}
\centering
\includegraphics[width=0.5\textwidth,clip=]{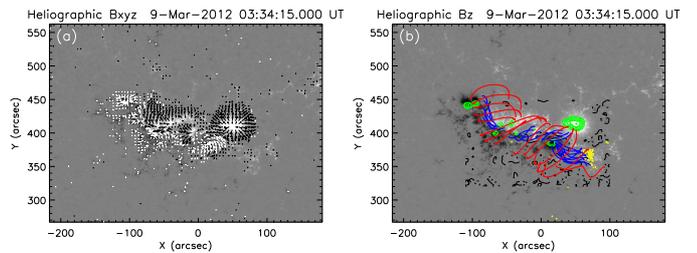}
\caption{(a) Projection-corrected vector magnetic field, where arrows represent the horizontal components.
(b) Radial magnetic field overlaid on the 3600 \AA \ contour (yellow and green lines), the polarity inversion line
 (dash-dotted lines), and the NLFFF field lines (blue and red lines). The magnetic field has been corrected
for projection effects. The yellow lines indicate the location of the WLF, while the green ones indicate the sunspot place.
} \label{fig3}
\end{figure}

Figure~3(a) shows the vector magnetic field observed by \textit{SDO}/HMI at 03:34 UT, which was 6 min before the WLF peak at 03:40 UT. The 180$^\circ$ ambiguity is removed by the minimum energy method \citep{Leka2009}. We first pre-process the vector magnetic field data to remove the magnetic force and torque \citep{Wiegelmann2006}. We then correct for projection effects and transform the line-of-sight and transverse components of the vector magnetic field to the heliographic components as described in \citet{Gary1990}. Finally, an NLFFF extrapolation is made to obtain the magnetic field configuration with the optimization method \citep{Wheatland2000, Wiegelmann2004}. Figure~3(b) depicts the magnetic field lines and the 3600~\AA \ contours. Highly sheared field lines with relatively low altitudes close to the polarity inversion line are clearly present below a set of higher magnetic arcades. Some field lines in the core field region are twisted around each other. The WLF is located at the western end of the sheared or twisted or both magnetic field lines, where the magnetic field is relatively weak.

\section{Discussion and conclusions}

A WLF was clearly observed on 2012 March 9 in different wavebands,
particularly in the 3600~\AA \ and 4250~\AA \ continuum bands by the new
telescope ONSET. Figure 2 clearly shows that the peak times of the emissions at
3.79 GHz, the H$\alpha$ line center, H$\alpha$ $6562.8 \pm 0.5$ \AA, 3600~\AA, 4250~\AA,
and the hard X-ray are temporally coincident. The contrast at
3600~\AA \ is larger than that at 4250~\AA, implying that the WLF has a Balmer
jump. These demonstrate that it is a type I WLF \citep{Fang1995}.
However, the general configurations in H$\alpha$
and 1700~\AA \ are quite different from that of the WLF: the first two have a ribbon-like structure, while the last has a point-like one, though both the H$\alpha$ and 1700~\AA \ images were relatively bright at both the WLF time and location.
This implies that what we observed at either 1700~\AA \ or in the H$\alpha$ wing
cannot represent the WL emission feature.

\begin{figure}
\centering
\includegraphics[width=0.39\textwidth,clip=]{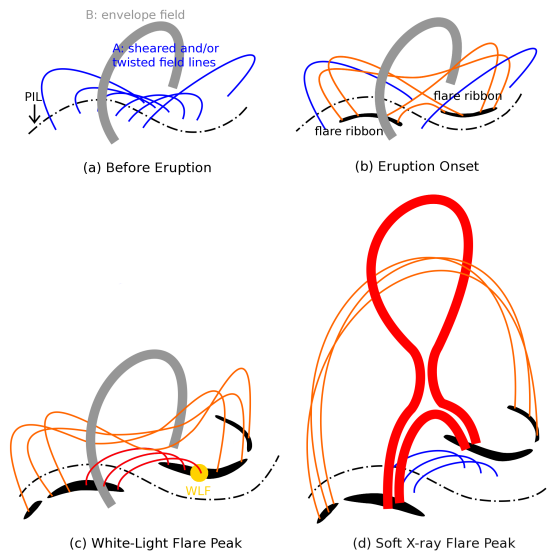}
\caption{A cartoon showing the two-step magnetic reconnection scenario, i.e., the first step
between the highly sheared or twisted field lines and the second in the current sheet stretched
by the erupting flux rope. Gray, blue, orange, and red colors indicate the different field line systems.
} \label{fig4}
\end{figure}

Using the vector magnetic field data from \textit{SDO}/HMI, we make an
NLFFF extrapolation to obtain the
magnetic field configuration around the WLF site and before the WLF peak time. Figure 3 shows that the WLF occurred at one end of the highly sheared or twisted field lines, which suggests that magnetic reconnection occurred
in the low-lying core field region. Otherwise, the energetic particles produced by
the magnetic reconnection cannot have precipitated to the footpoint of the core
field lines. A similar situation had been previously observed by \cite{Guo2012}, who
found that hard X-ray sources appear at the footpoints of an erupting
flux rope. Considering a microwave type III burst and hard
X-ray burst occurred all at the WLF time, we suggest that an
electron beam has accelerated and bombarded the lower atmosphere, producing
the WLF through collisional heating or backwarming.
The WLF only appeared in the region with positive polarity,
because the magnetic field there was relatively weak compared to the region of conjugate polarity
connected by the field line.

Figure~4 depicts a cartoon to show the entire mechanism and illustrate our
two-step magnetic reconnection scenario, i.e., magnetic reconnection
between the highly sheared or twisted field lines and that in the current
sheet stretched by the erupting flux rope. Before the magnetic reconnection
and the magnetic-flux rope eruption, the whole magnetic field configuration
is constituted by a core field with highly sheared or twisted field lines
and a potential envelope field before the flare onset as shown in Figure~4(a).
The core field has an curved elbow that is in the opposite direction at each end, which extends
out of the envelope field.

For the first-step magnetic reconnection, the core field first started to reconnect
as shown in Figure~4(b), which corresponds to the 1700~\AA \ and H$\alpha$ observations
at 03:25 UT. The 1700~\AA \ emission flashed at the center of the core field region,
and a helioseismic wave also appeared in the same region (J. W. Zhao, private communication).
Then, the inner-core field lines continued to reconnect with the outer, curved elbow
field-lines. The core field lines finally formed an erupting flux rope and the WLF appeared on the north side at the footpoints of the reconnected field lines below the reconnection
site at 03:40 UT, as shown in Figure~4(c). The microwave
peak at around 03:40 UT is lower than that at 03:56 UT. However, the former produced the
WLF, whereas the latter did not. This indicates that the first-step reconnection must occur at
a lower altitude than the second-step one. In this way, high energy particles generated in
the first-step reconnection can be relatively easily bombarded in the lower atmosphere.

For the second-step magnetic reconnection, the erupting flux rope stretches the envelope
field to form a current sheet. Magnetic reconnection in the current sheet produces the long
phase after the WLF peak. The main peak of the microwave burst appears at about 03:56 UT, when
there is no more WL emission excess. The H$\alpha$ ribbons clearly developed after 03:56 UT.
In particular, the soft X-ray peak appearing at 03:53 UT is 13 min later than the WLF time (around 03:40 UT).
All these phenomena can only be explained by the second-step magnetic reconnection. This two-step magnetic reconnection scenario is similar to the
tether-cutting model that has been applied to solar eruptions \citep{Moore2001}. We must point out
that the two-step reconnection is an approximate classification. However, it is already adequate to point out
the key differences between the magnetic reconnections in the core and envelope fields.

\begin{acknowledgements}
 We warmly thank the referee for advice that improved our paper in many ways.
 We also thank Dr. Yuying Liu, who acquired the radio data,
 Dr. Yang Liu, who provided the \textit{SDO}/HMI data, and Dr. Junwei Zhao,
 who provided the helioseismic wave information.
 This work was supported by the National Natural Science Foundation
 of China (NSFC) under the grant numbers 10878002, 10610099,
 10933003, 10673004 and 11103009, as well as a grant
 from the 973 project 2011CB811402.
\end{acknowledgements}


\end{document}